\documentclass[twocolumn,amsmath,amssymb,prl]{revtex4-1}
\usepackage{physics}
\usepackage[dvipdfmx]{graphicx} 
\usepackage{dcolumn} 
\usepackage{bm} 
\usepackage{braket} 
\usepackage[usenames,dvipsnames]{color}
\usepackage[colorlinks=true, citecolor=blue, urlcolor=blue, linkcolor=blue,
setpagesize=false, bookmarks=false
]{hyperref}
\usepackage{siunitx}
\usepackage{mhchem}

\begin{document}

\title{Wannier-Stark ladders and Stark shifts of excitons in Mott insulators}
\author{Mina Udono$^{1}$}
\author{Tatsuya Kaneko$^{2}$}
\author{Koudai Sugimoto$^{3,}$}
\email{sugimoto@rk.phys.keio.ac.jp}
\affiliation{$^1$Department of Physics, Chiba University, Chiba 263-8522, Japan\\
$^2$Department of Physics, Osaka University, Toyonaka, Osaka 560-0043, Japan\\
$^3$Department of Physics, Keio University, Yokohama, Kanagawa 223-8522, Japan}
\date{\today}

\begin{abstract}
External-field driven energy-level discretization, such as Landau quantization or Stark localization, is one of the most intriguing phenomena in quantum systems.
We investigate the emergence of the Wannier-Stark ladder coming from the particle-hole continuum and the Stark shifts of the exciton levels in one-dimensional Mott insulators under the dc electric field.
The discretized peak structure in the optical-conductivity spectra newly appears by applying the dc electric field, and the positions of these peaks can be reproduced from the energy levels of a simple effective model in the strong-coupling regime.
Our results not only suggest that Mott insulators can serve as a viable platform for Stark discretization, but also pave the way for investigations of dynamical properties in correlated many-body systems under a dc electric field.
\end{abstract}

\maketitle

The effects of external fields on quantum systems have been extensively studied for a long time, yet continue to present interesting issues.
The Stark effect \cite{STARK1913}, where the spectral lines of atoms and molecules split due to the presence of an external electric field, is one of the most famous phenomena.
In condensed-matter physics, electric-field effects predicted in seminal papers, e.g., Bloch oscillations \cite{Bloch1929}, Wannier-Stark localization \cite{PhysRev.117.432,PhysRevB.38.1667}, Zener tunneling \cite{doi:10.1098/rspa.1934.0116}, etc., have been demonstrated in semiconductors and their superlattices \cite{GLUCK2002103,WACKER20021,RevModPhys.90.021002}.
More recently, Stark band engineering and Stark shifts of excitons \cite{Chaves:2020aa} have been explored in van der Waals semiconductors, such as transition-metal dichalcogenides \cite{Ramasubramaniam2011,Klein:2016aa,Pedersen2016,Scharf2016,Leisgang:2020aa} and black phosphorus \cite{Liu:2017aa,Chaves2015}.
With regard to quantum many-body physics, Stark localization in disorder-free interacting systems attracts great attention due to its similarity to many-body localization \cite{schulz2019,nieuwenburg2019,morong2021}.
We also note that the dielectric breakdown of correlated systems by a strong dc electric field has been investigated both experimentally \cite{PhysRevB.62.7015,Guiot2013,Yamakawa2017} and theoretically \cite{PhysRevLett.91.066406,PhysRevLett.105.146404,PhysRevB.86.075148,PhysRevB.86.085127,Eckstein_2013,PhysRevB.89.205126,PhysRevB.98.075102}.

In Mott insulators (MIs), the Stark effects may prominently appear by applying the dc electric field when carriers, i.e., doublons (doubly occupied sites) and holons (empty sites), are created.
The energy continua of upper and lower Hubbard bands become discretized by the electric field and the energy spectrum forms the ladderlike structure similar to the Wannier-Stark ladder~\cite{Eckstein_2013,PhysRevB.89.205126,PhysRevB.98.075102}.
In addition, if there are doublon-holon interactions leading to form their bound states, i.e., excitons, the Stark shift of the exciton level may occur as in a single hydrogen atom \cite{PhysRevB.105.L241108}.
In this case, the energy spectrum is expected to acquire multiple structures derived from the simultaneous emergence of both the Wannier-Stark discretization of the Hubbard bands and the atomiclike Stark shift of the exciton level.

In this Letter, we investigate the optical conductivity of the one-dimensional extended Hubbard model under a dc electric field to clarify the energy-level discretization of MIs by employing the infinite time-evolving block decimation (iTEBD) method \cite{PhysRevLett.98.070201,Orus2008PRB}.
First, we demonstrate the appearance of the Wannier-Stark ladder by the discretization of the doublon-holon continuum in the optical conductivity.
In the strong-coupling regime, this discretization can be well reproduced by an effective model defined in a restricted subspace that permits the existence of only a single doublon and holon.
Second, by introducing intersite interactions, we show that the Stark shift of the exciton level below the Mott gap newly appears in addition to the Wannier-Stark ladder above the Mott gap.
The origins of these spectra are analyzed by the effective restricted-subspace model and the solvable two-site Hubbard model.
Lastly, we discuss the experimental feasibility of these effects using the parameters corresponding to a one-dimensional organic MI.

We consider the one-dimensional extended Hubbard model at half filling.
The Hamiltonian is given by
\begin{multline}
\hat{H}  =-t_{\mathrm{h}} \sum_{j,\sigma}(\hat{c}^{\dagger}_{j,\sigma}\hat{c}_{j+1,\sigma}+{\rm H.c.})\\
  +U\sum_{j}\hat{n}_{j,\uparrow}\hat{n}_{j,\downarrow}
+V\sum_{j}\hat{n}_{j}\hat{n}_{j+1},
\label{H}
\end{multline}
where $\hat{c}^{\dag}_{j,\sigma}$ ($\hat{c}_{j,\sigma}$) is the creation (annihilation) operator of a fermion at site $j$ with spin $\sigma$ ($=\uparrow,\downarrow$) and $\hat{n}_{j,\sigma} = \hat{c}^{\dag}_{j,\sigma} \hat{c}_{j,\sigma}$ ($\hat{n}_{j} = \hat{n}_{j,\uparrow} + \hat{n}_{j,\downarrow}$).
$t_{\mathrm{h}}$ is the hopping amplitude between the nearest-neighbor sites and is set as a unit of energy.
$U$ and $V$ are the on-site and nearest-neighbor repulsive interactions, respectively.
This model can capture the electronic properties of various one-dimensional MIs~\cite{PhysRevLett.81.657,Wall2011,PhysRevB.103.045124,PhysRevLett.101.177404}.

To calculate the optical conductivity numerically in the thermodynamic limit, we employ the iTEBD method~\cite{PhysRevLett.98.070201,Orus2008PRB}.
In our iTEBD calculations, we incorporate a spatially uniform electric field $E(t)$ by employing the Peierls substitution $t_{\mathrm{h}} \hat{c}^{\dagger}_{j,\sigma}\hat{c}_{j+1,\sigma} \rightarrow t_{\mathrm{h}} e^{-iqA(t)}\hat{c}^{\dagger}_{j,\sigma}\hat{c}_{j+1,\sigma}$, where $q$ represents the fermion charge and $A(t)$ is the vector potential satisfying $E(t) = -\partial_t A(t)$.
This approach is chosen over the alternative method where $H' = - q E(t) \sum_j R_j \hat{n}_{j}$, with $R_j$ being the position of site $j$, is incorporated into the Hamiltonian in Eq. \eqref{H} \cite{SciPostPhys.3.4.029}.
This is because the Peierls substitution is suitable for iTEBD calculations relying on the translational invariance.
Here we set the Planck constant and the lattice constant to $\hbar = 1$ and $a = 1$, respectively.
To obtain the quantum state under the dc electric field $E_0$, we first prepare the ground state without the electric field as an initial state at $t=0$, and then perform the numerical time evolution of the wave function with $A (t) = -\theta(t) E_0 t$, where $\theta (t)$ is a step function.

We estimate the optical conductivity from a response of a current by applying an additional weak electric field.
The current operator in a vector potential $A(t)$ is written as $\hat{J}_{A}(t) = iqt_{\mathrm{h}} \sum_{j,\sigma}(e^{iqA(t)}\hat{c}^{\dagger}_{j+1,\sigma}\hat{c}_{j,\sigma}-{\rm H.c.})$.
By applying additional infinitesimal electric field $\delta E(t)$ to the present steady state, the deviation of the current per site becomes $\delta j (t) = \frac{1}{L} \qty(\ev*{\hat{J}_{A + \delta A} (t)} - \ev*{\hat{J}_{A} (t)}) = \int^{t}_{-\infty} \sigma(t-t') \delta E(t') \dd{t'}$, where $L$ is the system size and $\sigma(t-t')$ represents the response function.
Then, the optical conductivity at the frequency $\omega$, which is the Fourier transform of $\sigma (t-t')$, is given by
\begin{align}
  \sigma(\omega)=\frac{\delta j (\omega)}{i(\omega+i\eta) \delta A(\omega)},
  \label{op}
\end{align}
where $\eta$ is a damping factor.
Although this factor is introduced for convergence of our numerical Fourier transformation, it is associated with the lifetime of the quasiparticles by, e.g., impurity scattering in actual materials. 
The method introduced here has also been used to calculate the nonequilibrium optical conductivity for the pump-probe spectroscopy~\cite{PhysRevB.93.195144,PhysRevB.98.165103,PhysRevB.91.245117,PhysRevB.104.085122}.
In our simulations, we adopt a weak probe pulse $\delta A(t) = A_{\rm pr}e^{-(t-t_{\rm pr})^2/2\sigma_{\rm pr}^2}{\rm cos}[\omega_{\rm pr}(t-t_{\rm pr})]$ in place of the infinitesimal external field.
We set $q=-1$, $A_{\rm pr}=0.01$, $\omega_{\rm pr}=10t_{\mathrm{h}}$, $\sigma_{\rm pr}=0.05/t_{\mathrm{h}}$, and $t_{\rm pr}=5/t_{\mathrm{h}}$.
Unless otherwise noted, we use $\eta = 0.1t_{\mathrm{h}}$.
We have numerically confirmed that the choice of $t_{\mathrm{pr}}$ ($> 0$) hardly affects our results, except for the last case discussed in this Letter.
In addition, we have also confirmed that the value of $A_{\mathrm{pr}}$ used is sufficiently small to obtain the linear response of the current.
Details of the numerical calculations are found in the Supplemental Material~\cite{SM}.

\begin{figure}[t]
\begin{center}
\includegraphics[width=\linewidth]{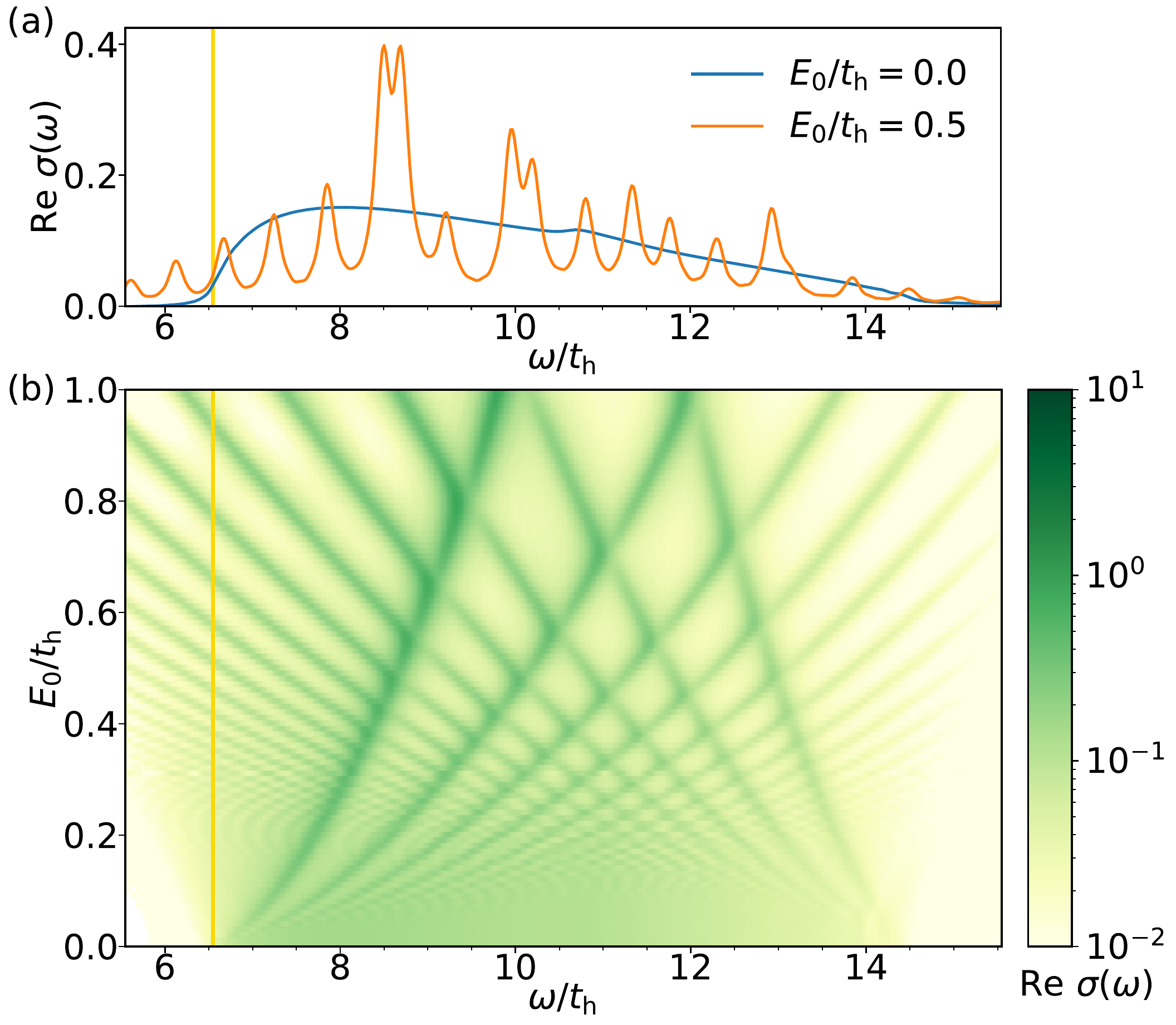}
\caption{
(a) 
$\Re \sigma (\omega)$ with and without the dc electric field $E_0$ at $U/t_{\mathrm{h}} = 10$ (and $V = 0$).
(b)
$\Re \sigma (\omega)$ in the plane of $\omega$ and $E_0$.
The vertical yellow line indicates the Mott gap evaluated by the Bethe ansatz \cite{essler2005one}.
}
\label{1}
\end{center}
\end{figure}

First, we present the results at $V=0$.
Figure~\ref{1}(a) shows the real part of the optical conductivity $\Re \sigma(\omega)$ at $U/t_{\mathrm{h}}=10$ with and without the electric field.
Note that the value of $E_0$ used in Fig.~\ref{1} is smaller than the threshold of dielectric breakdown~\cite{PhysRevB.86.075148}.
For the case of $E_0=0$, $\Re \sigma(\omega)$ exhibits a broad spectral weight originating from the doublon-holon continuum, which is present above the Mott gap $\Delta_{\rm M}$.
The width of this continuum is approximately $8t_{\mathrm{h}}$ \cite{PhysRevLett.85.3910,essler2005one}.
Upon applying the electric field, the broad spectral weight becomes discretized into multiple peaks.
To see this in more detail, we show the $E_0$ dependence of the optical spectra in Fig.~\ref{1}(b).
We find that the continuous spectrum at $E_0=0$ gradually gets discretized to multiple peaks as $E_0$ is increased, and at large $E_0$, the energies of the peak positions approach $\omega=U \pm mE_0$ with integer $m$, which implies the appearance of the Wannier-Stark ladder.
Note that the Wannier-Stark ladder appearing in the density of states of MIs has been investigated by dynamical mean-field theory \cite{Eckstein_2013,PhysRevB.89.205126,PhysRevB.98.075102}.

\begin{figure}[t]
\begin{center}
\includegraphics[width=\linewidth]{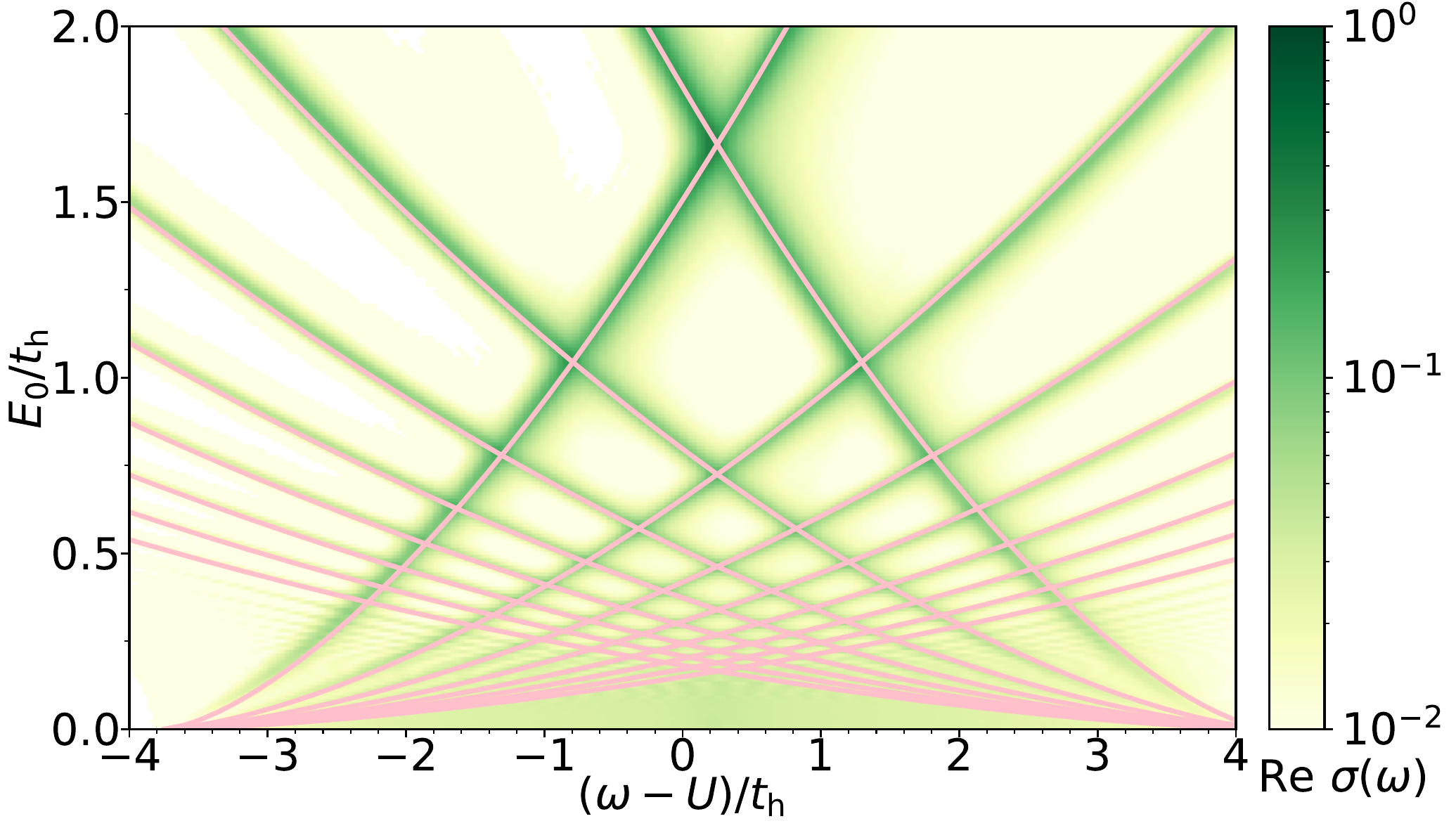}
\caption{
$\Re \sigma (\omega)$ in the plane of $\omega- U$ and $E_0$ at $U/t_{\mathrm{h}} = 40$ (and $V  = 0$).
The solid pink lines represent the energy spectra of $H_{\mathrm{WSL}}^{\pm}$ up to $\abs{R_{\mathrm{dh}}} = 100$, but the eight eigenvalues from the lowest (for $+$) and the highest (for $-$) are plotted.
Because of the many-body effects that are not incorporated in the simplified model $H_{\rm WSL}^{\pm}$, there is the slight difference in the center of energy with the iTEBD calculation.
To adjust this, the energy of $H_{\rm WSL}^{\pm}$ is shifted by $0.25t_{\mathrm{h}}$.
}
\label{2}
\end{center}
\end{figure}

In the strong-coupling limit ($U \gg t_{\mathrm{h}}, E_0$), the Wannier-Stark ladder emerging in the optical conductivity can be interpreted by employing a doublon-holon model in the restricted subspace where the excited states include only one doublon and one holon on a half-filled chain \cite{PhysRevB.56.15025}.
Here, we incorporate the dc electric field in the length gauge ($H' = - q E_0 \sum_j R_j \hat{n}_{j}$) because it is favorable for making a model in the real-space picture.
We define $\ket{R_{\mathrm{dh}}}$ as the state where the relative position of the doublon with respect to the holon is $R_{\mathrm{dh}}$.
Within this representation, the electric potential energy corresponding to the state $\ket{R_{\mathrm{dh}}}$ can be expressed as $-q E_0 R_{\mathrm{dh}}$.
By using this state as a basis, the effective Hamiltonian of the strong-coupling model with $q=-1$ becomes a tridiagonal matrix given by
\begin{align}
  H_{\rm WSL}^{\pm} =
  \left( \begin{array}{ccccc}
  U  \pm  E_0 & - 2t_{\mathrm{h}} &  0 & \cdots  & \\
  - 2t_{\mathrm{h}} & U \pm  2 E_0 & - 2t_{\mathrm{h}} & \ddots &   \\
  0 & - 2t_{\mathrm{h}} & U \pm  3 E_0 &  \ddots  \\
  \vdots & \ddots  & \ddots &\ddots\\
  \end{array} \right)
  ,
  \label{large}
\end{align}
where the sign $+$ ($-$) indicates that the basis states used are $\ket{R_{\mathrm{dh}} \geq +1}$ ($\ket{R_{\mathrm{dh}} \leq -1}$).
Each off-diagonal component is twice $-t_{\mathrm{h}}$ due to the presence of two equivalent processes that can change the state $\ket{R_{\mathrm{dh}}}$ to $\ket{R_{\mathrm{dh}} + 1}$: one where the doublon hops to the right site and the other where the holon hops to the left site.
In particular, at $E_0 \gg t_{\mathrm{h}}$, the energy spectrum is given by $\omega \simeq U \pm m E_0$, suggesting that the wave function is localized by the strong electrostatic field.
In Fig.~\ref{2}, we show the $E_0$ dependence of $\Re \sigma(\omega)$ at $U / t_{\mathrm{h}} = 40$ to see the behavior in the strong-coupling regime.
We find that the energy spectra of the effective model (solid lines) are in good agreement with the peak positions of $\Re \sigma(\omega)$.
Therefore, Eq.~\eqref{large} is valid for describing the Wannier-Stark ladder in the strong-coupling limit.

\begin{figure}[t]
\begin{center}
\includegraphics[width=\linewidth]{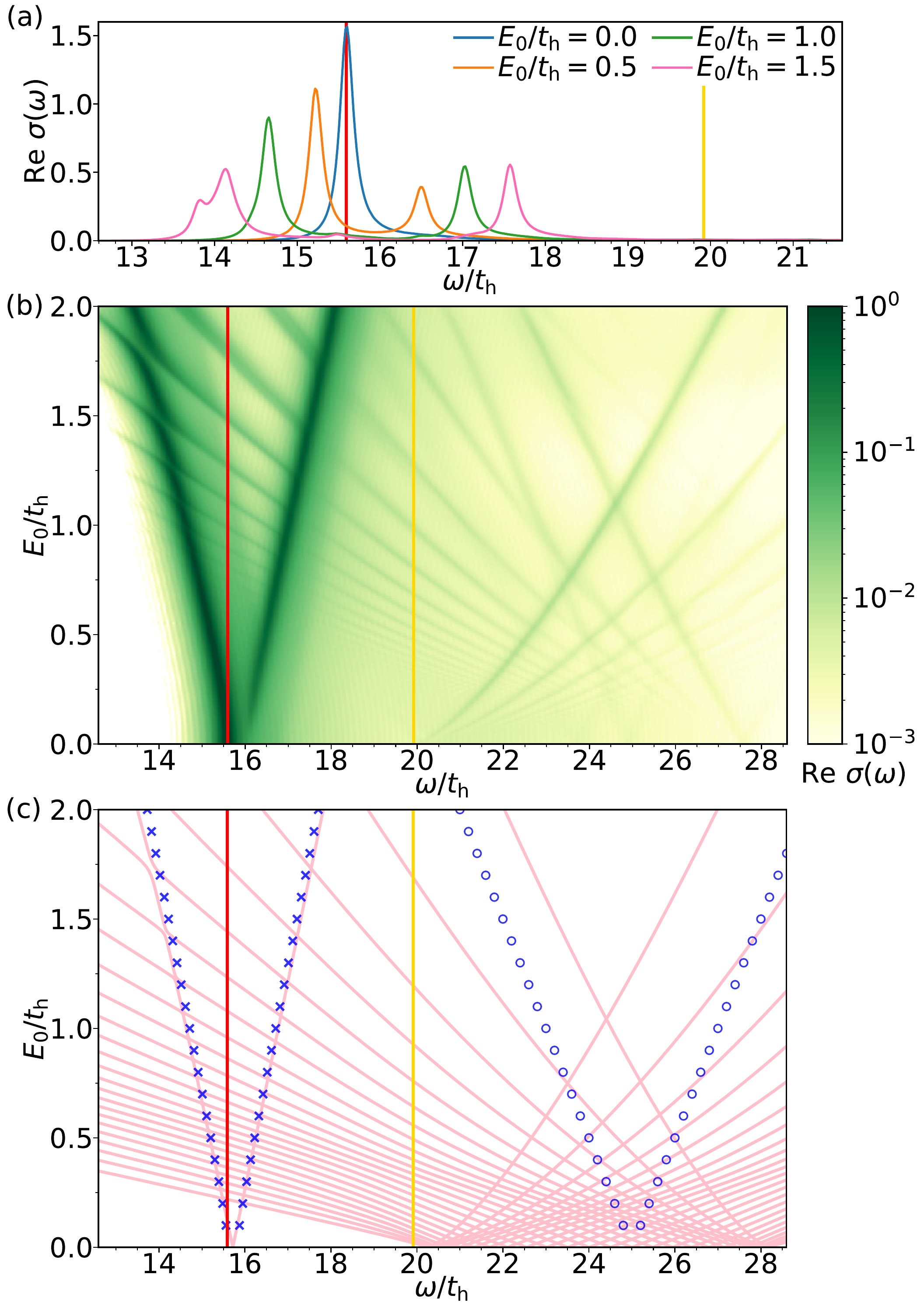}
\caption{
(a)
$\Re \sigma (\omega)$ at $U/t_{\mathrm{h}} = 24$ and $V /t_{\mathrm{h}} = 8$.
The red vertical line indicates the peak position of $\Re \sigma(\omega)$ at $E_0 = 0$, which we refer to as the (odd-parity) exciton level, and the yellow vertical line corresponds to the Mott-gap energy.
(b)
$\Re \sigma (\omega)$ in the plane of $\omega$ and $E_0$.
(c)
The energy spectra of $H_{\rm WSL}^{\pm}$ up to $\abs{R_{\mathrm{dh}}} = 30$, where the center of energy is shifted by $0.22t_{\mathrm{h}}$.
The blue cross marks are the energies of the two-site Hubbard model, where the energy is shifted by $-0.42t_{\mathrm{h}}$.
The blue circles represent the energies predicted by the biexciton model, where the energy is shifted by $t_{\mathrm{h}}$.
}
\label{3}
\end{center}
\end{figure}

Next, we consider the excitonic effects induced by the intersite interaction $V$.
In the one-dimensional extended Hubbard model, it is known that the exciton level becomes lower than the bottom of doublon-holon continuum when $V/t_{\mathrm{h}} > 2$ \cite{PhysRevB.54.R17269,PhysRevB.55.15368,PhysRevB.56.15025,PhysRevB.64.125119,PhysRevB.67.075106,PhysRevB.105.L241108}.
Figure~\ref{3}(a) shows $\Re \sigma(\omega)$ at $U/t_{\mathrm{h}}=24$ and $V/t_{\mathrm{h}} = 8$.
We choose the relatively large values of $U$ and $V$ to compare the iTEBD results with the strong-coupling model discussed later.
For $E_0 = 0$, in contrast to the case of $V=0$ where a broad spectrum appears above the Mott gap [see Fig.~\ref{1}(a)], the spectrum in the current situation displays a sharp peak below the Mott gap.
The energy of this peak corresponds to the exciton level, which arises due to the presence of the nonlocal interactions.
The red vertical lines in Fig.~\ref{3} indicate the energy of this peak.
While the excitations coming from the doublon-holon continuum still remain above the Mott gap \cite{PhysRevB.67.075106}, a large part of the spectral weight concentrates on this exciton peak.

When the electric field $E_0$ is applied, we observe the peak splitting, and the width of the split is proportional to $E_0$.
The split of the exciton level by the influence of the electric field resembles the Stark effect in a hydrogen atom resulting from the hybridizations of even- and odd-parity wave functions with respect to the static electric field \cite{Sakurai:1167961}.
In the extended Hubbard model, there are the odd- and even-parity doublon-holon bound states in the sub-Mott-gap regime \cite{PhysRevB.56.15025}.
Although the optical excitation to the even-parity exciton level is forbidden at $E_0=0$, the hybridization of the odd- and even-parity excitons by the dc field $E_0$ leads to two optically allowed exciton peaks with the linear Stark shift. 
We demonstrate this shift by the two-site Hubbard model under the strong electric field, which is valid since the doublon and holon are locally confined in the nearest-neighbor sites by large $V$.
Within this model, we obtain the energy levels of the excitons described by
$\epsilon_{\pm} = \qty(\epsilon^{\mathrm{(o)}}_{\mathrm{ex}} + \epsilon^{\mathrm{(e)}}_{\mathrm{ex}})/2 \pm \sqrt{\qty[\qty(\epsilon^{\mathrm{(o)}}_{\mathrm{ex}} - \epsilon^{\mathrm{(e)}}_{\mathrm{ex}})/2]^2 + F^2_{\mathrm{e}}}$,
where $\epsilon^{\mathrm{(o)}}_{\mathrm{ex}}$ and $\epsilon^{\mathrm{(e)}}_{\mathrm{ex}}$ represent the energy levels of the odd- and even-parity excitons, respectively, and $F_\mathrm{e}$ is a quantity proportional to the electric-field strength \cite{SM}.
Therefore, the split excitonic peaks appear in the optical conductivity under the strong electric field.

Figure \ref{3}(b) shows the detailed $E_0$ dependence of $\Re \sigma(\omega)$.
In addition to the peak structures around the exciton level, we observe the emergence of multiple peaks originating from the doublon-holon continuum above the Mott gap for $E_0 > 0$, which is analogous to the case of $V=0$.
These peaks can also be attributed to the Wannier-Stark discretization of the doublon-holon continuum.

These structures are reproduced through the diagonalization of the matrix of Eq.~\eqref{large}, replacing $U  \pm  E_0$ with $U - V \pm  E_0 $.
The solid pink lines in Fig.~\ref{3}(c) correspond to the energy of the strong-coupling model.
This model is qualitatively consistent with both the Wannier-Stark ladder and the split exciton peaks observed in Fig.~\ref{3}(b).
We find that the lower Wannier-Stark ladder belonging to the $U-mE_0$ ($m\ge 2$) sector enters into the energy region of the excitons.
While the lower exciton at $\omega = U-V-E_0$ hybridizes with the states belonging to the lower ladder, the upper exciton at $\omega = U-V+E_0$ does not couple with them due to differing doublon-holon configurations.
Note that while the energies of the even- and odd-parity excitons are not degenerate at $E_0=0$ \cite{PhysRevB.62.R4769}, the two exciton levels in the strong-coupling model based on Eq.~\eqref{large} are degenerate.
We also present the energy levels of the excitons determined from the two-site model as blue cross marks in Fig.~\ref{3}(c).
These marks are in good agreement with the exciton peaks presented in Fig.~\ref{3}(b).
Thus, the two-site model provides a better description of the Stark shift of the exciton compared to the model in Eq.~\eqref{large}.

Furthermore, we observe an additional peak structure along the line connecting $(\omega/t_{\mathrm{h}}, E_0 / t_{\mathrm{h}}) \sim (25,0)$ and $(20.5,2)$ in Fig.~{3}(b), which is absent in the energy spectra of $H^{\pm}_{\text{WSL}}$.
This structure is attributed to the Stark shift of a biexciton.
The energy levels of the odd- and even-parity biexcitons are given by $\epsilon^{\mathrm{(o)}}_{\mathrm{biex}} \simeq \epsilon^{\mathrm{(e)}}_{\mathrm{biex}} \simeq 2U-3V$ \cite{SM}.
The blue circles in Fig.~\ref{3}(c) denote the biexciton energies under the electric field, represented as $\epsilon_{\mathrm{biex},\pm} \simeq 2U - 3V \pm 2E_0$.
While the peak corresponding to $\epsilon_{\mathrm{biex},-}$ is visible, the peak corresponding to $\epsilon_{\mathrm{biex},+}$ is unfortunately too weak to be discernible in Fig.~\ref{3}(b).
We conclude that the energy spectrum in the strong-coupling case can be characterized by the Wannier-Stark ladder and Stark shifts of the excitons and biexcitons.

\begin{figure}[t]
\begin{center}
\includegraphics[width=\linewidth]{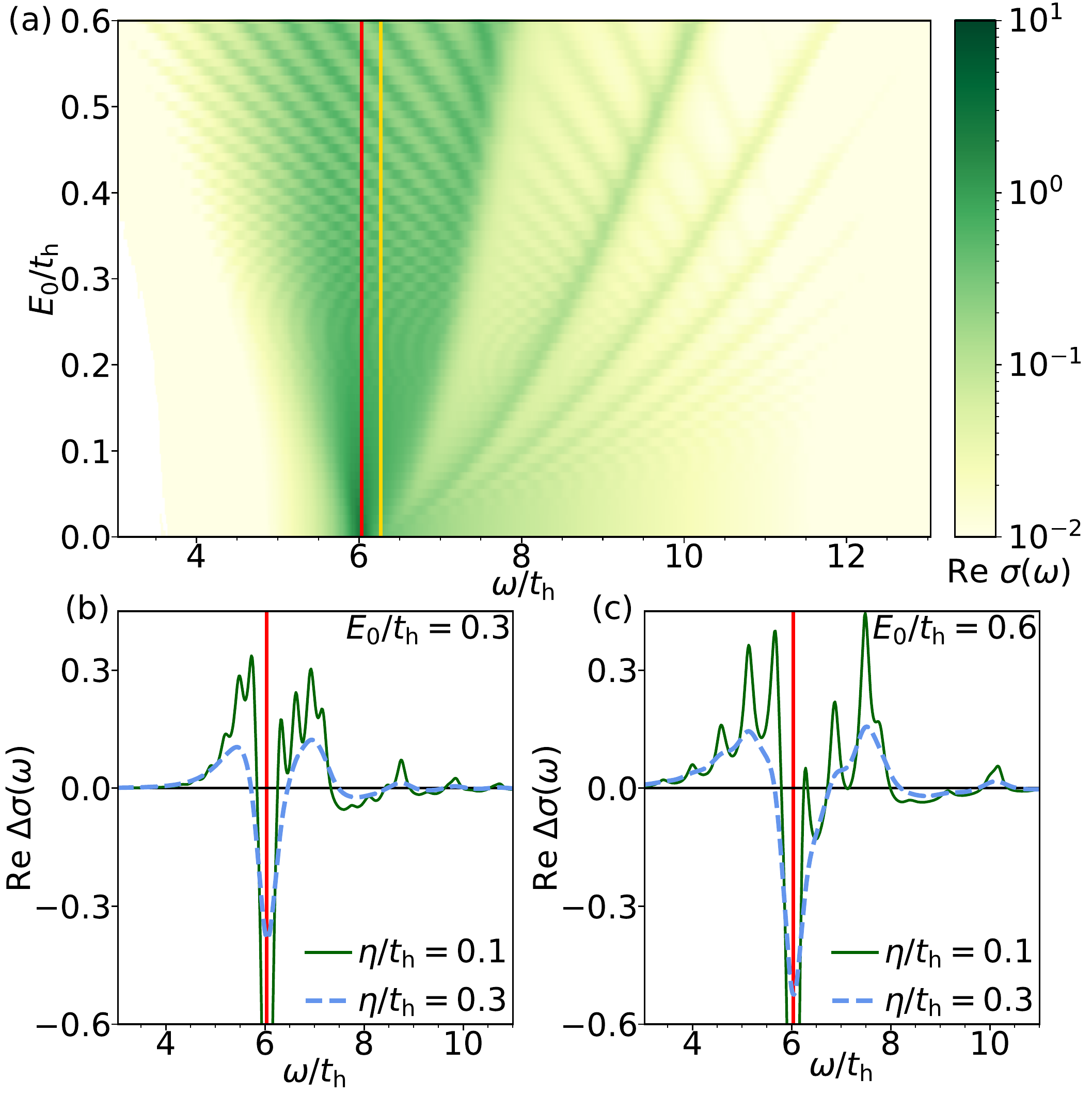}
\caption{
(a)
$\Re \sigma (\omega)$ at $U/t_{\mathrm{h}} = 10$ and $V /t_{\mathrm{h}} = 3$ in the plane of $\omega$ and $E_0$.
The vertical red and yellow lines indicate the exciton energy and the Mott gap, respectively.
(b),(c)
Change of the real part of optical conductivity $\Re \Delta\sigma(\omega)$ by varying the electric field from $E_0/t_{\mathrm{h}} = 0$ to (b) $E_0/t_{\mathrm{h}} = 0.3$ and (c) $E_0/t_{\mathrm{h}} = 0.6$.
The solid (dashed) lines are the results for the damping factor $\eta/t_{\mathrm{h}}=0.1$ ($\eta/t_{\mathrm{h}}=0.3$).
}
\label{4}
\end{center}
\end{figure}

Finally, we discuss the experimental feasibility of observing the many-body Stark effects. 
Figure~\ref{4}(a) shows $\Re \sigma(\omega)$ at $U/t_{\mathrm{h}} = 10$ and $V /t_{\mathrm{h}} = 3$, which corresponds to the parameters of the organic compound \ce{ET-F2TCNQ} \cite{PhysRevB.103.045124}.
In this calculation, we have found that the dc electric field should be introduced in a more adiabatic manner compared to the aforementioned cases to prevent excitations caused by the abrupt electric-field quench at $t=0$.
We use a vector potential with the form $A (t) = -\theta(t) E_0 [t - s\tanh (t/s)]$, which ensures $E(t) \simeq E_0$ for $t \gg s$.
Here, we set $s = 6 / t_{\mathrm{h}}$ and $t_{\mathrm{pr}}=30/ t_{\mathrm{h}}$.
Note that $E_0 / t_{\mathrm{h}} = 0.6$ in our calculation corresponds to $E_0 \sim 1.0$ \si{MV/cm} in \ce{ET-F2TCNQ} ($t_{\mathrm{h}} \sim 0.1$ \si{eV} and $a \sim 6$ \si{\AA}) \cite{Wall2011,PhysRevB.86.075148,PhysRevB.103.045124,HASEGAWA1997489}. 
Since the exciton level (red line) is close to the Mott gap (yellow line) in contrast to the case in Fig.~\ref{3}, the excitonic Stark splitting and Wannier-Stark ladder coming from the doublon-holon continuum are almost overlapped.
Although the Stark shifts of the exciton levels are not clear, the stripe structure due to the Wannier-Stark discretization still remains. 
Hence, even when using the parameters of an actual material, we can find the signature of Stark discretization. 
In experiments, the observability of Stark discretization of the optical spectra may strongly depend on the ratio between the magnitude of $E_0$ and the damping factor $\eta$.
If $E_0 \gg \eta$, the energy-level spacing is large enough, and well-separated multiple peaks are observable.
However, if $E_0 \sim \eta$, the individual peaks are smeared out \cite{SM}. 

While we use the dc electric field in our simulation, a similar situation can be made in experiments using the pump-probe technique if the pump pulse frequency $\Omega$ is sufficiently smaller than the gap, such as the terahertz (\ce{THz}) range~\cite{Schmidt:2018aa}.
THz pump-probe spectroscopy to \ce{ET-F2TCNQ} has been reported in Ref.~\cite{Miyamoto2019}, where the electric-field-induced changes of the optical spectra exhibit the plus-minus-plus structure around the exciton energy.
As shown in Figs.~\ref{4}(b) and \ref{4}(c), our calculation can reproduce a similar plus-minus-plus structure in the field-induced change of the real part of the optical conductivity $\Re \Delta \sigma(\omega)$ when $\eta$ is comparable to $E_0$.
Besides this, the suppression of $\eta$ unveils the multiple peaks due to Stark discretization [see Figs.~\ref{4}(b) and \ref{4}(c)].
Hence, Stark discretized peaks emerge in pump-probe spectra if damping effects are suppressed and/or a strong THz pump field is applied. 

In summary, we have revealed the energy spectra of the MI under the dc electric ﬁelds by calculating the optical conductivities.
The spectra show the Wannier-Stark ladder emerging from the doublon-holon continuum and the Stark shift of the exciton level.
These energy levels in the strong-coupling regime are well reproduced by the simple effective models for the Wannier-Stark ladder and excitons. 
Moreover, we have demonstrated the effect using the parameters corresponding to the one-dimensional MI, \ce{ET-F2TCNQ}, and have suggested a pathway to experimental realization.

\begin{acknowledgments}
The authors would like to thank Y. Ohta and M. Sato for their valuable comments.
This work was supported by Grants-in-Aid for Scientific Research from JSPS (Grants No. JP18K13509, No. JP19K14644, No. JP20H01849, No. JP21K03439, and No. JP23K03286).
M.U. acknowledges the support by JST, the establishment of university fellowships towards the creation of science technology innovation (Grant No. JPMJFS2107).
The iTEBD and density-matrix renormalization-group calculations were performed using the ITensor library~\cite{itensor}.
\end{acknowledgments}

\bibliography{References}

\end{document}


\title{Supplemental Material:\\ Wannier-Stark ladders and Stark shifts of excitons in Mott insulators}
\author{Mina Udono$^{1}$}
\author{Tatsuya Kaneko$^{2}$}
\author{Koudai Sugimoto$^{3}$}
\affiliation{
$^1$Department of Physics, Chiba University, Chiba 263-8522, Japan\\
$^2$Department of Physics, Osaka University, Toyonaka, Osaka 560-0043, Japan\\
$^3$Department of Physics, Keio University, Yokohama, Kanagawa 223-8522, Japan
}
\date{\today}

\maketitle

\renewcommand\thefigure{S\arabic{figure}}
\setcounter{figure}{0}
\renewcommand{\bibnumfmt}[1]{[S#1]}
\renewcommand{\citenumfont}[1]{S#1}
\renewcommand{\theequation}{S\arabic{equation}}

\section{Details of numerical calculations}

In this study, we employ the iTEBD method \cite{PhysRevLett.98.070201,Orus2008PRB} for the calculation of the ground state and its time evolution.
In Fig.~1(a) of the main text, calculations were performed with a bond dimension of $\chi = 3000$, but for the rest, we confirmed that the results sufficiently converge with $\chi = 2000$.
For time evolution, we set the time step $\delta t = 0.01$ and use the fourth-order Trotter decomposition.

As mentioned in the main text, when we calculate the optical conductivity numerically, we estimate a response function by the time evolution of an electric current induced by a weak probe light.
Strictly speaking, the precise optical conductivity should be evaluated from the Kubo formula with infinite-boundary conditions (IBC)~\cite{Phien2012PRB, Phien2013PRB, Zauner2015JPCM, PhysRevB.105.L241108}.
However, the computational cost for calculating the current-current correlation function far exceeds that of the time-dependent simulation of the electric current.
It is worth noting that we have calculated the optical conductivity using IBC with several parameters and confirmed that the results obtained by the probe-light method are almost identical to the results from the Kubo formula.

The Mott gaps at $V \neq 0$ are estimated by the finite-size density-matrix renormalization group method~\cite{White1992PRL, White1993PRB, Schollwock201196}.
The Mott gap is obtained by calculating $\Delta_{\rm M} (L) = E_{\rm gs}(L, L+1) + E_{\rm gs}(L, L-1) - 2 E_{\rm gs}(L,L)$, where $E_{\rm gs}(L,N)$ is the ground-state energy for an $L$-site system with $N$ particles.
We numerically calculate $\Delta_{\rm M} (L)$ up to $L=256$ with the periodic-boundary conditions, and perform the extrapolation to the thermodynamic limit based on the $L$ dependence.

\section{Two-site Hubbard model}

The excitons considered in this study are formed locally by the interactions between the nearest-neighbor sites.
In other words, we expect that the Hubbard model with only two sites can help understand the behavior of excitons under electric fields.
In the following, we will discuss energy levels of excitons and the Stark shift under the dc electric field.

The two-site Hubbard model is given by
\begin{equation}
  \hat{H}_{\text{2-site}} \!
    = \! -t_{\rm h}
      \sum_{\sigma} \qty(\hat{c}^{\dagger}_{1,\sigma} \hat{c}_{2,\sigma} \!+\! \mathrm{H.c.})
      + U \sum_{j=1}^2 \hat{n}_{j,\uparrow} \hat{n}_{j,\downarrow}
      + V \hat{n}_{1} \hat{n}_{2}.
\end{equation}
We consider the set of states belonging to $S_z=0$ sector at half-filling, i.e., $\ket{\uparrow,\downarrow} = \hat{c}^{\dag}_{1,\uparrow}\hat{c}^{\dag}_{2,\downarrow} \ket{0}$,
$\ket{\downarrow,\uparrow}=\hat{c}^{\dag}_{1,\downarrow}\hat{c}^{\dag}_{2,\uparrow}\ket{0}$,  $\ket{\uparrow\downarrow,0}=\hat{c}^{\dag}_{1,\uparrow}\hat{c}^{\dag}_{1,\downarrow}\ket{0}$, and $\ket{0,\uparrow\downarrow}=\hat{c}^{\dag}_{2,\uparrow}\hat{c}^{\dag}_{2,\downarrow}\ket{0}$.
It is convenient to choose the symmetry-adapted basis written as
\begin{align}
  \ket{\psi_0^{\rm (s)}}
    &=\frac{1}{\sqrt{2}} \qty[\ket{\uparrow,\downarrow}-\ket{\downarrow,\uparrow}], \\
  \ket{\psi_0^{\rm (t)}}
    &=\frac{1}{\sqrt{2}} \qty[\ket{\uparrow,\downarrow}+\ket{\downarrow,\uparrow}], \\
  \ket{\psi_0^{\rm (e)}}
    &=\frac{1}{\sqrt{2}} \qty[\ket{\uparrow\downarrow,0}+\ket{0,\uparrow\downarrow}],
\intertext{and}
  \ket{\psi_0^{\rm (o)}}
    &=\frac{1}{\sqrt{2}} \qty[\ket{\uparrow\downarrow,0}-\ket{0,\uparrow\downarrow}].
\end{align}
$\ket{\psi_0^{\rm (s)}}$ and $\ket{\psi_0^{\rm (t)}}$ are spin singlet and triplet states, respectively. 
Besides, $\ket{\psi_0^{\rm (e)}}$ and $\ket{\psi_0^{\rm (o)}}$ are the even and odd spacial parity doublon-holon states, respectively. 
Note that the spin-triplet state $\ket{\psi_0^{\rm (t)}}$ is irrelevant in the following discussion because its total spin is different from the ground state and the applied electric fields do not affect the sector of this basis.
Using the three remaining basis, $\ket{\psi_0^{\rm (s)}}$, $\ket{\psi_0^{\rm (e)}}$, and $\ket{\psi_0^{\rm (o)}}$, the matrix representation of $\hat{H}_{\text{2-site}}$ is given by
\begin{align}
\begin{split}
H_{\text{2-site}}
=\left( \begin{array}{ccc}
		0 & \gamma & 0 \\
		\gamma & U-V & 0 \\
		0 & 0 & U-V \\
	\end{array} \right),
\end{split}
\label{eq:Hmat}
\end{align}
where $\gamma = -2t_{\mathrm{h}}$ $(<0)$.
Note that we subtract the constant term (such as  $V$) in the Hamiltonian matrix and do the same hereafter.
Since the parity of $\ket{\psi_0^{\rm (o)}}$ is different from the others (i.e., the off-diagonal elements are zero), the state corresponding to the odd-parity exciton is just given by
\begin{align}
\ket{\psi_{\rm ex}^{\rm (o)}}= \ket{\psi_0^{\rm (o)}}
\end{align}
with energy
\begin{align}
&\epsilon_{\rm ex}^{\rm (o)}=U-V.
\end{align}
Note that the energy of odd-parity exciton of the one-dimensional extended Hubbard chain is given by $U-V-4t_{\mathrm{h}}^2/V$ in the strong-coupling limit \cite{PhysRevB.56.15025}.
On the other hand, $\ket{\psi_0^{\rm (s)}}$ and $\ket{\psi_0^{\rm (e)}}$ are hybridized by the hopping integral. 
Hence, diagonalization of the $2\times 2$ matrix in (\ref{eq:Hmat}) gives the ground state $\ket{\psi_{\rm g}^{\rm (s)}}$ and the even-parity exciton state written as
\begin{align}
  \ket{\psi_{\rm ex}^{\rm (e)}}
    \!=\! \sqrt{\frac{1}{2} \qty(1 \!+\! \frac{U \!-\! V}{D})} \ket{\psi_0^{\rm (e)}}
   \! - \! \sqrt{\frac{1}{2} \qty(1 \!-\! \frac{U \!-\! V}{D})} \ket{\psi_0^{\rm (s)}}, 
\end{align}
where $D=\sqrt{\qty(U-V)^2 + 4\gamma^2}$ and its energy is
\begin{align}
  \epsilon_{\rm ex}^{\rm (e)}
    = \frac{U - V + D}{2}.
\end{align}
At $U-V \gg t_{\mathrm{h}}$, the energy reads
\begin{align}
\epsilon_{\rm ex}^{\rm (e)} \simeq U - V + \dfrac{4t_{\mathrm{h}}^2}{U-V}. 
\end{align}

The Hamiltonian of the external dc electric field is given by $\hat{H}_E = - qE_0\sum_{j=1}^2 R_j \hat{n}_{j}$.
Under this electric field, the matrix representation of $\hat{H}_{\text{2-site}} + \hat{H}_E$ becomes
\begin{align}
\begin{pmatrix}
  0 & \gamma & 0 \\
  \gamma & U-V & F \\
  0 & F & U-V
\end{pmatrix},
\label{eq:Hmat_F}
\end{align}
with $F= qE_0$.
The energies obtained from this matrix are plotted in Fig.~3(c) of the main text with blue-cross marks.

The explicit form of the Stark shift can be estimated at $U-V \gg t_{\mathrm{h}}$ and $E_0 \gg t_{\mathrm{h}}$.
Using $\ket{\psi_{\rm g}^{\rm (s)}}$, $\ket{\psi_{\rm ex}^{\rm (e)}}$, and $\ket{\psi_{\rm ex}^{\rm (o)}}$ as a set of the basis of the Hamiltonian, the electric-field effect can be described by the matrix 
\begin{align}
\begin{pmatrix}
  \epsilon_{\rm g} & 0 & F_{\rm g} \\
  0 & \epsilon_{\rm ex}^{\rm (e)} & F_{\rm e}  \\
  F_{\rm g} & F_{\rm e} & \epsilon_{\rm ex}^{\rm (o)}
\end{pmatrix}
\end{align}
with $F_{\rm e} \propto E_0 \sqrt{\tfrac{1}{2}[1+(U-V)/D]}$ and $F_{\rm g} \propto E_0 \sqrt{\tfrac{1}{2}[1-(U-V)/D]}$.
Since $F_{\rm g}$ is small when $U-V \gg t_{\mathrm{h}}$, we assume the exciton-state sector is decoupled from the ground-state sector in the Hamiltonian matrix.
Within this approximation, the energies of the exciton levels are
\begin{equation}
  \epsilon_{\pm}
    = \frac{\epsilon_{\rm ex}^{\rm (e)}+\epsilon_{\rm ex}^{\rm (o)}}{2}
    \pm \sqrt{\qty(\frac{\epsilon_{\rm ex}^{\rm (e)}-\epsilon_{\rm ex}^{\rm (o)}}{2})^2 + F^2_{\rm e}},
\end{equation} 
and thus we obtain the approximate energy splitting of the Stark shift as
\begin{align}
\epsilon_{+} - \epsilon_{-}
\simeq 2F_{\rm e} + \frac{\left( \epsilon_{\rm ex}^{\rm (e)}-\epsilon_{\rm ex}^{\rm (o)} \right)^2}{4F_{\rm e}}
.
\end{align}

\section{Stark shift of biexciton}
It is known that the energy level of the biexciton in the extended Hubbard model is located at $2U-3V$ in the strong coupling limit \cite{PhysRevB.67.075106}.
Neglecting contributions from $t_{\mathrm{h}}$ and considering the even and odd-parity biexciton states as a set of basis, the matrix representation of the Hamiltonian under the electric field is given by
\begin{align}
	\left( \begin{array}{cc}
		\epsilon_{\rm biex}^{\rm (e)} & F_{\rm biex} \\
		F_{\rm biex} & \epsilon_{\rm biex}^{\rm (o)} \\
	\end{array} \right),
\label{stark}
\end{align}
where we assume $\epsilon_{\rm biex}^{\rm (o)}=\epsilon_{\rm biex}^{\rm (e)}=2U-3V$ and $F_{\rm biex}=2F=2qE_0$.
The factor of 2 in $F_{\rm biex}$ arises because the biexciton consists of two excitons.
From this matrix, we obtain the energies $\epsilon_{\rm biex,\pm} = 2U-3V \pm 2E_0$.
These energies at $U/t_{\mathrm{h}}=24$ and $V/t_{\mathrm{h}}=8$ are plotted in Fig.~3(c) of the main text as blue circles, which successfully reproduce the observed trend of the additional peak structure shown in Fig.~3(b) of the main text.

\section{Effect of damping factor $\eta$}

\begin{figure}[b]
\begin{center}
\includegraphics[width=\columnwidth]{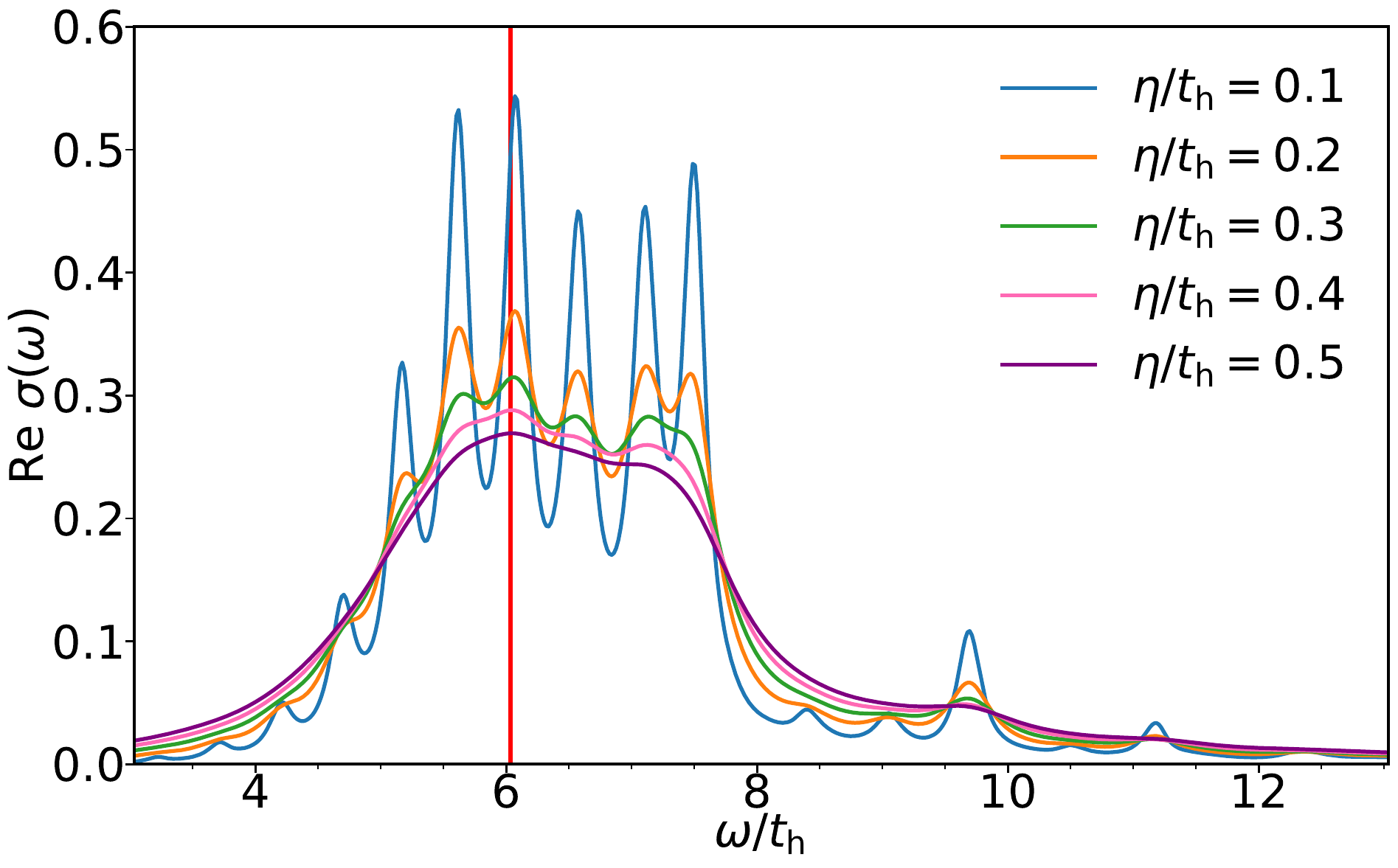}
\caption{
$\Re \sigma(\omega)$ for the extended Hubbard model with $U/t_{\mathrm{h}} = 10$ and $V/t_{\mathrm{h}} = 3$ under $E_0/t_{\mathrm{h}} = 0.5$ for various values of $\eta$.
The vertical red line indicates the position of the exciton energy.
}
\label{s1}
\end{center}
\end{figure}

In our numerical calculations, the damping factor remains an arbitrary value.
The damping factor, which is necessary for the convergence of the Fourier transformation, inevitably appears in actual experimental spectra due to the lifetime of quasiparticles.
We show $\Re \sigma(\omega)$ for different values of $\eta$ in Fig.~\ref{s1}.
We find that the discretized individual peaks are smeared out at $\eta/t_{\mathrm{h}} > 0.3$, forming a plateau-like broad structure around the exciton energy.
To observe the Wannier-Stark ladder in experiments, a sample with a damping factor smaller than the interval of the spectral peaks is required, or an electric field that creates a sufficiently large interval between the spectral peak is needed.

\bibliography{supplemental_material.bbl}